\documentclass[twocolumn,letterpaper,secnumarabic,amssymb, nobibnotes, aps, prl]{revtex4-1}        
\usepackage{epsfig}
\usepackage{dcolumn}
\usepackage{epstopdf}
\usepackage{graphicx}
\begin{document}

\title{Macroscopic magnetization in uniform magnetic fields}  

 \author{S. Selenu}
\affiliation{}

\begin{abstract}
\noindent  The finding of a new  formulation  of the magnetization vector of  a quantum system interacting with a static uniform magnetic field\cite{Selenu1} is reported. There  a  gauge invariant form of its divergence is shown being expressed as a function of the electronic  current per state coupled with the Berry curvature of the quantum system. A Fourier analysis of the magnetization vector and magnetization density is reported as an application of the presented formula it could be applied in the context of computational modelling\cite{Martin} of quantum matter.
\end{abstract}

\date{\today} 
\maketitle

 \section{Introduction}
\label{hg}
An ab initio framework  at the nano and micro scale here is adopted where projecting of electronic circuits becomes nowaday demanding. In fact, since the introduction in \cite{Selenu1,Resta,Vanderbilt,Resta1} of the quantum magnetization vector\cite{Jackson} of a quantal body it has been put an effort onto the research of the nature of the magnetization itself and its cause in order to understand how to control within the context of nano and micro technology its magnitude. Therefore it has also put out the need of explaining how the latter is related with the measured macroscopic electric current flowing through the magnetized body and why the latter being solenoidal since now avoided either classically by Maxwell formulation of the classical electromagnetism\cite{Jackson}. Notwithstanding  it is put on evidence main features of the proposed model with the scope of reaching a comprehensive theory of quantum electromagnetism at its first stage of its developing. In the next sections it will be shown how to formulate an expression of the quantum magnetization written in terms of the electronic current per state and in the last section how to find its source in terms of a topological invariant density related to the  Berry curvature of the quantum system, where meaning of the latter  was avoided since its discovery. Conclusions will be reported at the end of the article.

\section{Quantum topological source of Magnetization}

Here it will be reported a model of the quantum magnetization\cite{Selenu1}$\bf{M}$  expressed in terms of the macroscopic electric current carried by the electronic field of a quantum body interacting with a macroscopic static uniform magnetic field. The former written as its contribution per state:

\begin{eqnarray}
\label{equazionemag}
{\bf{M_n}}= \frac{e}{mc}\langle \Psi_{n,{\bf{k}}}|{\bf{p}\times i\nabla_{{\bf{k}} }}|\Psi_{n,{\bf{k}}}\rangle 
\end{eqnarray}

where $\bf{p}$ is the quantum momentum operator, and  $i\nabla_{{\bf{k}}}$ is the operator making varying the wave vectors  $\bf{k}$ of the electric wave $\Psi_n$ in its $n$-th state , allows  to calculate directly magnetization mean values of  $\bf{M}$ as:

\begin{eqnarray}
\label{equazioneMT}
 {\bf{M}}=\frac{1}{V}\sum_n{ f_n \bf{M_n}}
\end{eqnarray}
where $V$ is the volume of the body and $f_n$ the occupation numbers. The expression reported in (\ref{equazioneMT}) allows to express further the latter in terms of a macroscopic current carried by electrons of charge $e$ by taking into account the fact that the partial derivative  of any quantum eigenstate $|\Psi_n\rangle$ is :

\begin{eqnarray}
\label{equazione}
 | \partial_{\lambda}\Psi\rangle= \langle \Psi| \partial_{\lambda}| \Psi \rangle |\Psi\rangle 
\end{eqnarray}

allowing to reformulate equation (\ref{equazioneMT}) into a more useful form in order to show magnetization dependence on the macroscopic electric current per state. Let us proceed further by  firstly substituting eq.(\ref{equazione}) in eq.(\ref{equazioneMT}) where is the case of calculating the partial derivative with respect to the wave vector $\bf{k}$:

\begin{eqnarray}
\label{equazionemagjxx}
{\bf{M_n}}= \frac{e}{mc}\langle \Psi_{n,{\bf{k}}}|{\bf{p}}|\Psi_{n,{\bf{k}}}\rangle \times \langle \Psi_{n,{\bf{k}}}|i\nabla_{{\bf{k}}}|\Psi_{n,{\bf{k}}}\rangle 
\end{eqnarray}

The summing  over $n$ vibrational states of the electronic wave allow reaching a satisfactory result, that of having magnetization represented as a function of the electronic current contribution, whose macroscopic weighted average is:

\begin{eqnarray}
\label{equazionemagjxxMT}
{\bf{M}} = \frac{e}{mc}\frac{1}{V}\sum_n f_n \langle \Psi_{n,{\bf{k}}}|{\bf{p}}|\Psi_{n,{\bf{k}}}\rangle \times \langle \Psi_{n,{\bf{k}}}|i\nabla_{{\bf{k}}}|\Psi_{n,{\bf{k}}}\rangle 
\end{eqnarray}

Evaluating the divergence of the magnetization in the wave vector  space $\bf{k}$  directly shows its dependence on the quantum topological invariant per state called Berry curvature of the  quantum system, in fact we can calculate a quantum magnetization density  $\phi({\bf{k}})=\nabla_{{\bf{k}}}\cdot {\bf{M}}$ as the following expression, where we avoid the label $\bf{k}$ on wave functions for the sake of notation:

\begin{eqnarray}
\label{equazioneDM}
 \phi({\bf{k}})&&= \frac{e}{mc}\frac{1}{V}\sum_n f_n \nabla_{{\bf{k}}}\cdot [\langle \Psi_{n}|{\bf{p}}|\Psi_{n}\rangle \times \langle \Psi_{n}|i\nabla_{\bf{k}}|\Psi_{n} \rangle] \\\nonumber
&& =\frac{1}{V}\sum_n f_n [\frac{1}{c}{\bf{J}}_{n} \cdot [\nabla_{{\bf{k}}} \times \langle \Psi_{n}|i\nabla_{{\bf{k}}}|\Psi_{n} \rangle] \\\nonumber
&& =\frac{i}{V}\sum_n f_n [\frac{1}{c}{\bf{J}}_{n}\cdot [{\bf{rot }}\langle \Psi_{n}|\nabla_{{\bf{k}}}|\Psi_{n}\rangle] \\\nonumber
\end{eqnarray}

when is the case of a constant transport number $N_{n}=\langle \Psi_{n}|\Psi_{n}\rangle$, being also writing the $n$ state current as it follows:

\begin{eqnarray}
\label{equazioneJn}
{\bf{J}}_{n} = \langle \Psi_{n,{\bf{k}}}|{\bf{p}}|\Psi_{n,{\bf{k}}}\rangle \\\nonumber
\end{eqnarray}

In order to apply computationally via ab initio model schemes from first principles the found  formula  of the magnetization vector or either its divergenceit can be performed a Fourier analysis by firstly writing the quantum electronic wave as a Fourier expansion on wave vectors ${\bf{k}}$ then calculate formulas found until on a discrete mesh of wave vectors we shall call ${\bf{G}}$ in order to quantify the latter. Let us define the electronic wave as:

\begin{eqnarray}
\label{equazion}
\Psi &&=\int \Psi_{{\bf{k}}}d{\bf{k}} \\\nonumber
N&&=\int d{\bf{r}} \Psi^* \Psi=\int {N_{\bf{k}}}d{\bf{k}} 
\end{eqnarray}

being calling $N$ the total transport number of the electronic wave and ${N_{\bf{k}}}$ the continuous transport number dependent on $\bf{k}$ vectors. Let us also consider as a  discretization, a discrete mesh of $\bf{G}$ vectors then  write:

\begin{eqnarray}
\label{equazionh}
\Psi ({\bf{r}})= \sum_{\bf{G}} c_{\bf{G}}e^{i \bf{G}\cdot{r}} 
\end{eqnarray}

usually called plane wave expansion in common ab initio models\cite{Martin}, to which corresponds an electronic current per state equal to 
${\bf{J}}_{\bf{G}} =  |c_{\bf{G}}|^2 \hbar \bf{G}$ then reducing the magnetization formula to:

\begin{eqnarray}
\label{M}
{\bf{M}} = \frac{1}{V}\sum_{\bf{G}} f_{\bf{G}}  [\frac{e}{mc}\ |c_{\bf{G}}|^2 \hbar {\bf{G}}\times  c^*_{\bf{G}}i\nabla c_{\bf{G}}] 
\end{eqnarray}

reaching a discretized formula in Fourier space that can be applied in any ab initio modelling at the scope of   projecting quantum mechanically  magnetization vector of either nano or micro electronic circuits. Moreover,  by making considerations on regard to the explanation of the quantum nature of magnetization, it is shown the latter  being directly related to the Berry curvature of the system appearing useful then to express  in Fourier space  also the quantum magnetic density whose formula is:

\begin{eqnarray}
\label{Mh}
\phi(\bf{k}) &&= \frac{i}{V}\sum_{\bf{G}} f_{\bf{G}} [\frac{e}{mc}|c_{\bf{G}}|^2 \hbar \bf{G} \cdot [\nabla c^*_{\bf{G}} \times \nabla c_{\bf{G}}] 
\end{eqnarray}

recognizing then how to quantify the topological invariants in terms of well defined measurable electronic waves amplitude and associated wave vectors, it put on the position of concerning future engeneering of the found formulas by mean of Fourier analysis of electronic signals at the nano and microscale. In the next part of the article it will be reported conclusions.

\section{Conclusions}
A formula of the  magnetization vector of a quantum system  of electrons is reported either in a continuous or a discrete formalism  where the found ab initio modelling allows  to quantify magnetic behaviour of  a quantum system  in its magnetic states.  Also, it is shown its strict relation with a Berry curvature  on the expression of its density put then into evidence the latter being   quantum topological invariants amenable of  a quantum engeneering via a Fourier analysis of experimental data.


\noindent 






\begin{thebibliography}{0} \expandafter\ifx\csname
natexlab\endcsname\relax\def\natexlab#1{#1}\fi \expandafter\ifx\csname
bibnamefont\endcsname\relax \def\bibnamefont#1{#1}\fi \expandafter\ifx\csname
bibfnamefont\endcsname\relax \def\bibfnamefont#1{#1}\fi
\expandafter\ifx\csname citenamefont\endcsname\relax
\def\citenamefont#1{#1}\fi \expandafter\ifx\csname url\endcsname\relax
\def\url#1{\texttt{#1}}\fi \expandafter\ifx\csname
urlprefix\endcsname\relax\def\urlprefix{URL }\fi
\providecommand{\bibinfo}[2]{#2} \providecommand{\eprint}[2][]{\url{#2}}




\bibitem{Selenu1}S.Selenu,  https://arxiv.org/abs/2011.14082
\bibitem{Martin}R.M. Martin,Electronic structure(Cambridge University Press, 2004)

\bibitem{Resta}R. Resta, Davide Ceresoli, T. Thonhauser, David Vanderbilt, Orbital Magnetization in Extended Systems,
First published: 05 September 2005 https://doi.org/10.1002/cphc.200400641

\bibitem{Vanderbilt} Davide Ceresoli,T. Thonhauser,David Vanderbilt,and R. Resta,  PHYSICAL REVIEW B74, 024408 (2006)

\bibitem{Resta1}R.Resta,  Magnetic circular dichroism versus orbital magnetization, PHYSICAL REVIEW RESEARCH 2, 023139 (2020)
\bibitem{Jackson}J.D. Jackson, Classical Electrodynamics(John wiley andSons, 1975)

\bibitem{Bloch}Felix Bloch (1928). "Über die Quantenmechanik der Elektronen in Kristallgittern". Zeitschrift für Physik (in German). 52 (7–8): 555–600. Bibcode:1929ZPhy...52..555B. doi:10.1007/BF01339455. S2CID 120668259.
\bibitem{Kittel} C. Kittel ,Introduction to Solid State Physics(John wiley$\&$ Sons.,Inc., NewYork, Chichester, Brisbane, Toronto,Singapore, 1996).
\bibitem{Ascroft}N. W. Ashcroft and N. D. Mermin,Solid State Physics(Saunders College Publishing, 1976)
\bibitem{Landau1}L.D Landau, E.M. Lifshitz Quantum Mechanics(Perga-mon Press London - Paris, 1958).

\end{thebibliography}
\end{document}